%% file: main.tex
\documentclass[final, journal,compsoc]{IEEEtran}
\IEEEoverridecommandlockouts

\usepackage{comment}
\usepackage{float}
\usepackage[binary-units]{siunitx}
\usepackage[ruled,vlined,linesnumbered]{algorithm2e}
\usepackage{multirow}
\usepackage{soul}
\usepackage{graphicx}
\usepackage{booktabs}
\usepackage{amsmath}
\usepackage{amssymb}
\usepackage{amsthm}
\usepackage{breqn}
\usepackage{soul,color}
\usepackage{makecell}
\newcommand{\bhl}[1]{{#1}}
\newcommand{\cb}{}

\begin{document}

\title{CloudProphet: A Machine Learning-Based Performance Prediction for Public Clouds}

\author{Darong~Huang,
        Luis~Costero,
        Ali~Pahlevan,
        Marina~Zapater,~\IEEEmembership{Member,~IEEE,}
        David~Atienza,~\IEEEmembership{Fellow,~IEEE}%
\thanks{Darong Huang, Ali Pahlevan, and David Atienza are with the Embedded Systems Laboratory (ESL), École polytechnique fédérale de Lausanne (EPFL), 1015 Lausanne, Switzerland. 
E-mail: \{darong.huang, ali.pahlevan, david.atienza\}@epfl.ch.

Luis Costero was with the Embedded System Laboratory (ESL), École Polytechnique Fédérale de Lausanne (EPFL), 1015 Lausanne, Switzerland. Now, he is with the Dpto. of Computer Architecture and Automatics, Complutense University of Madrid (UCM), Spain (email: lcostero@ucm.es).

Marina Zapater was with the Embedded System Laboratory (ESL), École Polytechnique Fédérale de Lausanne (EPFL), 1015 Lausanne, Switzerland. Now, she is with the School of Engineering and Management of Vaud (HEIG-VD), University of Applied Sciences Western Switzerland (HES-SO), 1401 Yverdon-les-Bains, Switzerland. (e-mail: marina.zapater@heig-vd.ch).
}
}

\IEEEtitleabstractindextext{ %
\begin{abstract}
Computing servers have played a key role in developing and processing emerging compute-intensive applications in recent years. Consolidating multiple virtual machines (VMs) inside one server to run various applications introduces severe competence for limited resources among VMs. Many techniques such as VM scheduling and resource provisioning are proposed to maximize the cost-efficiency of the computing servers while alleviating the performance inference between VMs. However, these management techniques require accurate performance prediction of the application running inside the VM, which is challenging to get in the public cloud due to the black-box nature of the VMs.
From this perspective, this paper proposes a novel machine learning-based performance prediction approach for applications running in the cloud. To achieve high accuracy predictions for black-box VMs, the proposed method first identifies the running application inside the virtual machine. It then selects highly-correlated runtime metrics as the input of the machine learning approach to accurately predict the performance level of the cloud application.
Experimental results with state-of-the-art cloud benchmarks demonstrate that our proposed method outperforms the existing prediction methods by more than 2x in terms of worst prediction error.
In addition, we successfully tackle the challenge in performance prediction for applications with variable workloads by introducing the performance degradation index, which other comparison methods fail to consider.
The workflow versatility of the proposed approach has been verified with different modern servers and VM configurations.
\end{abstract}
    
}

\maketitle

\input{s1_intro}
\input{s2_related_work}
\input{s3_problem_description}
\input{s5_proposed_method}

\input{s6_exp_setup}

\input{s7_exp_results}

\section{\bhl{Conclusions and Future Work}}
\label{sec:conclusion}
In this paper, we have proposed a machine learning-based performance prediction method for cloud applications based on the realistic assumption that resource governors should regard VMs as black-box systems. The proposed method first identifies the application based on accessible hardware metrics from the host server. Next, highly-correlated metrics are selected to accurately predict the performance level of the application by using the proposed machine learning-based approach. Finally, the performance degradation is predicted for the VM, which can be further used by the resource governor to schedule or migrate the VM.

Experiments with state-of-the-art cloud benchmarks demonstrate that our proposed machine learning-based method outperforms the state-of-the-art comparison approaches. We observed that the proposed method can achieve improvements of more than 2x in terms of worst prediction error. In addition, the proposed method performs accurate performance degradation prediction for applications with the variable workload while the comparison methods fail to consider.

In addition, we also showed that the proposed method can be ported to new server and VM configurations without degradation in prediction accuracy. When studying the trade-off between sampling time and prediction accuracy of the proposed method, the main conclusion is that the proposed method can achieve high prediction accuracy with a reasonable sampling time for the application.

Finally, the computation overhead was analyzed in detail when compared to the state-of-the-art methods. Although the proposed method has the largest computation overhead due to its highly sophisticated structure and high accuracy, the overhead is negligible when compared to the application execution time (microseconds vs. minutes/hours).

\bhl{In terms of future endeavors, several key aspects need attention and improvement:

Integration: Our forthcoming efforts will concentrate on achieving a profound integration of CloudProphet, our machine learning-based performance degradation method, within the public cloud monitoring and management system. This integration aims to facilitate real-time monitoring of performance degradation of VMs and ensure seamless operation.

Cost Savings: To determine the actual benefits of CloudProphet for both cloud providers and customers, we intend to perform assessments utilizing real public clouds. This evaluation will enable us to fully explore the advantages of incorporating artificial intelligence into cloud computing, highlighting its potential benefits.

Nevertheless, several challenges persist in our future endeavors that demand further consideration and resolution for AI and Cloud systems~\cite{gill2022ai}:

Automation: A significant challenge lies in integrating the machine learning-based performance degradation prediction method into the cloud monitoring and management system, with particular emphasis on devising strategies to effectively manage running virtual machines (VMs). Developing an automated approach to this process remains a critical obstacle that needs to be addressed.

Inadequate Data and Privacy Concerns: An ongoing concern revolves around the availability of sufficient data for performance monitoring and prediction, as well as the associated privacy issues. Determining the boundaries between the data that the public cloud provider can collect and the data users are willing to share requires careful examination. Consequently, it is imperative to explore potential solutions that address data scarcity and privacy concerns while maintaining the efficacy of machine learning-based performance degradation prediction methods.}

\section*{Acknowledgment}
This work has been partially supported by the EC H2020 RECIPE FET-HPC project (No. 801137), the ERC Consolidator Grant COMPUSAPIEN (No. 725657), an Industrial Grant from Huawei Cloud to ESL-EPFL, and Grant PID2021-126576NB-I00 funded by MCIN/AEI/10.13039/501100011033 and by "ERDF A way of making Europe".

\bibliographystyle{IEEEtran}
\bibliography{biblio}

\input{bio}

\end{document}

%% file: s1_intro.tex
\section{Introduction}

\IEEEPARstart{C}{loud} platforms have gained tremendous growth in the last decades because of their vast advantages in security, flexibility, and cost-efficiency~\cite{cortez_2017_resourcecentral}. Therefore, attracting end-users to transition their applications to the cloud.  \bhl{In the post-COVID future, cloud-computing investment increased by 37\% in the first quarter of 2020~\cite{gill2022ai}.} With this trend, worldwide end-user spending on cloud services is forecast to gain over 20\% average growth in recent years and reach around 400 billion dollars in 2022~\cite{garter_forecast}.

\begin{figure}[t]
    \centering
    \includegraphics[width=\columnwidth]{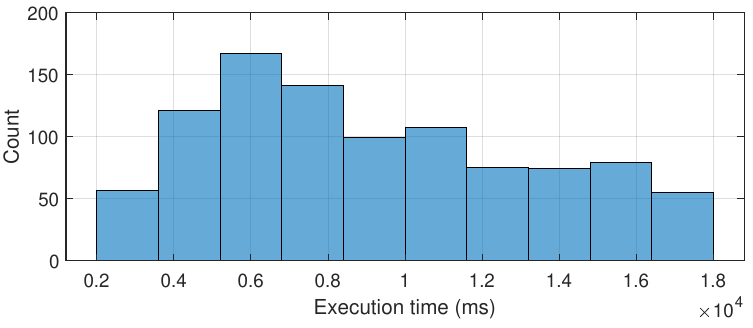}
    \caption{Performance variation of the Data Serving benchmark~\cite{cloudsuite} experienced on the cloud while collocating with multiple VMs. The results are obtained on a physical server by following the experiment setting introduced in Section~\ref{sec:exp_setup}}
    \label{fig:ds_perf_dist}
\end{figure}

The demand for public clouds has increased drastically in the last decade, entailing an explosion in energy usage. Data centers consume roughly 200 terawatt-hours of energy each year, which contributes to 1\% of global electricity demand~\cite{jones2018information}. The demand keeps ramping up, and estimations show that data centers will use around 6-10\% of global electricity in 2030~\cite{andrae2015global}. This motivates cloud service providers, particularly Amazon, Microsoft, Google, and Huawei, among others, to optimize the efficiency and usage of cloud servers towards a more sustainable and economical way~\cite{cortez_2017_resourcecentral, gao_2014_mlofordatacenter}. Thanks to the virtualization technology supported by off-the-shelf many-core server microprocessors, such as Intel VT and AMD-V~\cite{neiger_2006_intelvt,amd_v}, cloud providers can consolidate multiple independent virtual machines (VMs) in a single physical server to exploit the performance potential and energy efficiency of the server~\cite{cortez_2017_resourcecentral}.
Although virtualization technologies can guarantee resources isolation between VMs,
such as dedicated computing cores, memory size, and disk space between different users or applications, existing virtualization technologies cannot ensure performance isolation between VMs inside a single computing node. VMs still need to compete with each other for shared computing resources, such as last-level cache (LLC), memory and disk bandwidth, etc. Consequently, competition in these shared resources will degrade VMs' performance dramatically~\cite{kim_2013_vmconsolidation}.
In real cloud servers, cloud providers cannot access VMs created by their clients and read the performance metric of the running application due to privacy policies. The black-box nature of VMs restricts the cloud providers' ability to estimate the runtime status of VMs and host servers. In addition, this dilemma usually leads to an unoptimized VMs organization and imposes non-negligible performance degradation on VMs. Fig.~\ref{fig:ds_perf_dist} demonstrates the performance variation of the Data Serving benchmark~\cite{cloudsuite} running on the cloud collocated with multiple VMs. The execution time of Data Serving when receiving severe interference from other VMs is eight times longer than the best energy efficiency and shortest execution time it can achieve.

To alleviate the performance degradation induced by the VM collocation, 
both heuristic and machine learning-based VM management methods have been proposed in recent years~\cite{kim_2013_vmconsolidation, anwar_2019_vmmigration, akbar_2019_vmallocation}, which rely on the runtime performance of the VM to guide the VM's allocation and migration management.
Despite these prior efforts, runtime performance prediction of the VM, the essential metric that VM's management methods rely upon, has been inadequately studied.
Previous approaches tackle this problem by either assuming the performance level of the application is known to the resource governor~\cite{jin_2015_TCC_VMcollocation} or assuming the resource governor knows which application is running inside the VM, applying then a performance inference model to predict the performance level of the application~\cite{kim_2013_vmconsolidation}.
However, these assumptions cannot be applied to public clouds as all the running VMs need to be regarded as black boxes. Indeed, cloud providers cannot ``peek inside" a VM to gather usage statistics~\cite{wood2009sandpiper}. For instance, Amazon Elastic Computing Cloud and Huawei Elastic Cloud servers provide ``barebone" VMs where clients load their own operating system images and cloud providers cannot intervene.
In this regard, performance prediction remains a challenge because the only available information is low-level hardware metrics monitored from outside the VM, i.e., the host server.

Another challenge of performance prediction is the difficulty in predicting the performance degradation level, while the application has varying workloads.
For such applications, both the interference received from other VMs or a heavier workload can lead to a degraded performance level (e.g., longer execution time or larger latency). Consequently, this increases the difficulty in performance prediction.
\bhl{However, existing performance prediction schemes~\cite{Wang_2016_spark_interference, pham_2017_TCC_executiontimepredict, shekhar_2018_interference_predict, bader2022lotaru} only focus on performance prediction without interference and fail to consider the performance degradation induced by resource contention.}
Without discerning the effect of the interference or varying workload on the performance metric, the performance predictor from the existing works will give wrong predictions. Thus misleading the resource governor to apply unreasonable actions, finally introducing unnecessary management overhead and performance degradation.

In summary, current VM management techniques still lack a state-of-the-art performance prediction method to predict the performance level of VMs in dynamic workload scenarios and guide the resource governor to allocate and migrate VMs for lower performance degradation. In this context, we propose CloudProphet, a machine learning-based performance prediction for public clouds. More specifically, the contributions of this work are listed as follows:
\bhl{
\begin{itemize}
    \item We propose an application type identification method based on the dynamic time warping algorithm. The proposed application type identification method can accurately identify which application type is running inside the VM only with hardware counters information observed from the host server instead of the VM.
    \item We propose a neural network (NN)-based VM performance prediction method. The proposed method uses highly correlated metrics as input and predicts the performance level of the application accurately.
    \item On the basis of the proposed NN-based performance prediction method, we enable the prediction of performance degradation for applications in dynamic scenarios by discerning whether the performance variation is caused by workload variation or interference from other VMs.
    \item Experiments demonstrate that our proposed method has over 90\% accuracy in application type identification and performance prediction for cloud benchmarks. 
    \item We compare our proposed method against state-of-the-art approaches, obtaining improvements of more than 2x in terms of worst prediction error. In addition, we achieve accurate performance degradation prediction with variable workload applications, which the comparison methods fail to consider.
    Moreover, we prove the validity of our approach with different server and VM configurations.
\end{itemize}
}

The rest of this paper is organized as follows. Section~\ref{sec:related_work} reviews related work. In Section~\ref{sec:system_and_problem}, we provide an overview of the system characterization of the server as well as the problem description. In Section~\ref{sec:proposed_method}, we introduce our proposed machine learning-based performance prediction method. Sections~\ref{sec:exp_setup} and~\ref{sec:exp_results} present the experimental setup and results, followed by the conclusions in Section~\ref{sec:conclusion}.

%% file: s2_related_work.tex
\section{Related Work}
\label{sec:related_work}

As illustrated in the Introduction, the performance prediction method for VMs is critical in optimizing the usage of the computing server while lowering the performance degradation induced by collocated VMs. There are two common ways widely studied in the literature to provide such information: (1) future workload prediction and (2) runtime performance prediction, which will be introduced in this section.
\subsection{Future workload prediction}

Given the predicted near-feature workload for the VMs, such as CPU utilization, memory occupation, etc., dynamic resource management methods can proactively allocate or migrate VMs, thus adapting to the workload variation and lowering the performance interference between allocated VMs. Numerous works have focused on workload prediction, and various prediction technologies have been utilized. For instance, a linear regression model is proposed in~\cite{yang_2013_regression_workload} to estimate the incoming workload and then scale the cloud configuration to adapt to the future workload.
In addition to regression-based methods, 
random forest-based workload prediction methods are presented in~\cite{cetinski_2015_random_tree_workload, cao_2018_dbworkload}. In~\cite{zhong_2018_svm_workload}, a support vector machine-based workload prediction method is implemented to predict the workloads of the host server.

To further improve the prediction accuracy, more machine learning (ML) -based workload prediction methods have been presented in~\cite{gao_2020_mlworkload,Jayakumar_2020_genericworkload}. In particular, a cluster-based workload prediction method is proposed in~\cite{gao_2020_mlworkload}, which first clusters tasks into several categories and then predicts the utilization of CPU and memory
for task scheduling. Another work tackling the prediction problem using the long-short-term-memory technique to provide workload prediction is described in~\cite{Jayakumar_2020_genericworkload}.

However, highly dynamic and random requests of clients entail a tremendous challenge in predicting cloud servers' workload. The best prediction error is still 18\% according to~\cite{Jayakumar_2020_genericworkload}. Therefore, unreliable workload prediction schemes can lead resource governors to make non-optimal decisions, such as under-provisioning, or over-provisioning for VM management~\cite{masdari_2020_workloadsurvey}.

\subsection{Runtime performance prediction}

In addition to managing VMs based on predicted workload levels, resource governors can also manage VMs relying on their runtime performance level.
To identify the performance level of collocated VMs inside a data center, heuristic performance prediction methods are studied in~\cite{novakovic_2013_deepdive,vasic_2012_dejavu}. By comparing the performance level of the VM in the public cloud with the performance of a cloned VM running inside a dedicated sandbox environment, the performance level of the VM can be identified~\cite{novakovic_2013_deepdive,vasic_2012_dejavu}. However, this method suffers from the high computation cost of holding a sandbox environment and difficulties in measuring the performance level of the VM.

To avoid the large overhead of building a sandbox environment, Wang \textit{et al.}~\cite{Wang_2016_spark_interference} developed an analytical model to estimate performance interference among multiple Apache Spark tasks running concurrently on a computing server. Then, the model can be used in runtime to predict the performance of the Apache Spark application. As for the latency-sensitive applications, an ML-based method is proposed to predict system performance at runtime~\cite{shekhar_2018_interference_predict}. However, these methods~\cite{Wang_2016_spark_interference,shekhar_2018_interference_predict} are solely designed for the Spark application and lack the support for other cloud applications.

To estimate the performance interference for a wide range of cloud applications, Kim \textit{et al.}~\cite{kim_2013_vmconsolidation} targeted some iconic applications and enumerated all of the available application combinations to establish a detailed performance interference profile for each application combination. However, this method formulates the interference estimation as a combinational problem and makes it impossible to solve with a large number of VMs and applications.
To quickly and accurately identify incoming applications and estimate possible interference they will receive, a collaborative filtering technique is proposed in~\cite{delimitrou_2013_paragon}.
This approach first analyzes the incoming application to schedule them on the specific server, thus achieving the lowest interference among applications while the highest utilization of the server.
However, this method can only take effect at the phase of the deploying application, i.e., the proposed method fails to predict performance and take actions after the deployment at runtime.

There are also several works focusing on predicting the application's performance at runtime. Examples of these works are Aspen~\cite{spafford_2012_aspen}, Palm~\cite{tallent_2014_palm}, PEMOGEN~\cite{bhattacharyya_2014_pemogen}, and COMPASS~\cite{lee_2015_compass}. They all make the prediction based on the knowledge of the source code and the detailed runtime state of the applications, e.g., the number of instructions executed on the VM. Although these methods can give relatively accurate results, the requirement of source code availability and execution state make these methods impossible to apply on public cloud platforms due to the black-box nature of the VMs. To get over the limitation of these previous studies, a two-stage ML-based workflow execution time prediction method is proposed by Pham \textit{et al.}~\cite{pham_2017_TCC_executiontimepredict}. In this research, various runtime information (i.e., application types, VM configurations, and cloud server platforms) is collected and analyzed to predict the execution time of the specific application. However, execution time is the sole performance metric considered in~\cite{pham_2017_TCC_executiontimepredict}, and the interference among VMs is not considered.

\bhl{In summary, Table~\ref{table:comparison} compares the important key parameters of the related works with our proposed method, CloudProphet. The table presents a side-by-side comparison of various aspects, such as black-box scenario assumption, prediction target, technique used, and experiment settings.} To the best of our knowledge, there is still no comprehensive work considering performance prediction for cloud applications in a black-box system. More specifically, no work has been proposed to accurately predict the performance of applications with variable workload levels. Therefore, in this work, we present a novel ML-based performance prediction method for cloud applications to tackle existing challenges.

\begin{table*}[t]
\centering
\small
\caption{\bhl{Comparison of related work}}
\label{table:libvirt_statistics}

\cb

\scalebox{0.9}{
    \begin{tabular}{ c c c c c c c c c} 
    \toprule
    \textbf{Approach} & \textbf{black-box} & \textbf{ Prediction Target} & \textbf{Technique} & \textbf{Experiments} \\
    \midrule
    Yang \textit{et al.}~\cite{yang_2013_regression_workload} & $\times$  & Workload & Linear regression & CloudSim simulation \\
    \midrule
    Cetinski \textit{et al.}~\cite{cetinski_2015_random_tree_workload} & $\times$  & Workload & Random Forest & AuverGrid dataset  \\
    \midrule
    Cao \textit{et al.}~\cite{cao_2018_dbworkload} & $\times$ & Workload & Machine learning based classification & Real servers\\
    \midrule
    Zhong \textit{et al.}~\cite{zhong_2018_svm_workload} & $\times$ & Workload & Wavelet SVM & Google cloud dataset \\
    \midrule
    Gao \textit{et al.}~\cite{gao_2020_mlworkload} & $\times$  & Workload & Clustering & Google cloud dataset \\
    \midrule
    Jayakumar \textit{et al.}~\cite{Jayakumar_2020_genericworkload} & $\times$  & Workload & LSTM & Public datasets \\
    \midrule
    Novakovic \textit{et al.}~\cite{novakovic_2013_deepdive} & $\times$  &  Performance & Heuristic performance model & Real servers \\
    \midrule
    Vasic \textit{et al.}~\cite{vasic_2012_dejavu} & $\times$   &  Performance & Heuristic performance model & Real servers \\
    \midrule
    Shekhar \textit{et al.}~\cite{shekhar_2018_interference_predict} & $\times$  &  Performance & Gaussian Processes-based machine learning & Real servers \\
    \midrule
    Kim \textit{et al.}~\cite{kim_2013_vmconsolidation} & $\times$  &  Performance & Heuristic performance model & Real servers \\
    \midrule
    Delimitrou \textit{et al.}~\cite{delimitrou_2013_paragon} & $\times$  &  Performance & Collaborative filtering & Real servers \\
    \midrule
    Spafford \textit{et al.}~\cite{spafford_2012_aspen} & $\times$  &  Performance & Source-code based performance modeling & Real servers \\
    \midrule
    Tallent \textit{et al.}~\cite{tallent_2014_palm} & $\times$ &  Performance & Source code annotation & Real servers \\
    \midrule
    Bhattacharyya \textit{et al.}~\cite{bhattacharyya_2014_pemogen} & $\times$ &  Performance & Runitme performance model generation & Real servers \\
    \midrule
    Lee \textit{et al.}~\cite{lee_2015_compass} & $\times$  &  Performance & Automated static analysis & Real servers \\
    \midrule
    Pham \textit{et al.}~\cite{pham_2017_TCC_executiontimepredict} & $\times$  &  Performance & Random forest & Real servers \\
    \midrule
   \makecell{This work\\ (CloudProphet)} & \checkmark   &  \makecell{Workload-aware\\ performance degradation} & Machine learning based method & Real servers \\
    \bottomrule
    \end{tabular}
    }
    \label{table:comparison}
\end{table*}

%% file: s3_problem_description.tex
\section{System Characterization and Problem Description}
\label{sec:system_and_problem}

\subsection{Could computing and virtualization}
This work focuses on the Infrastructure as a Service (IaaS) cloud, which is widely provided by commercial public cloud service platforms, such as Amazon, Google, Microsoft, Huawei, etc. Thanks to the virtualization technology, the user can create multiple virtual machines
and share computing resources at the same time with other users without knowing the underlying hardware management and interactions.

\begin{figure}[t]
    \centering
    \includegraphics[width=\columnwidth]{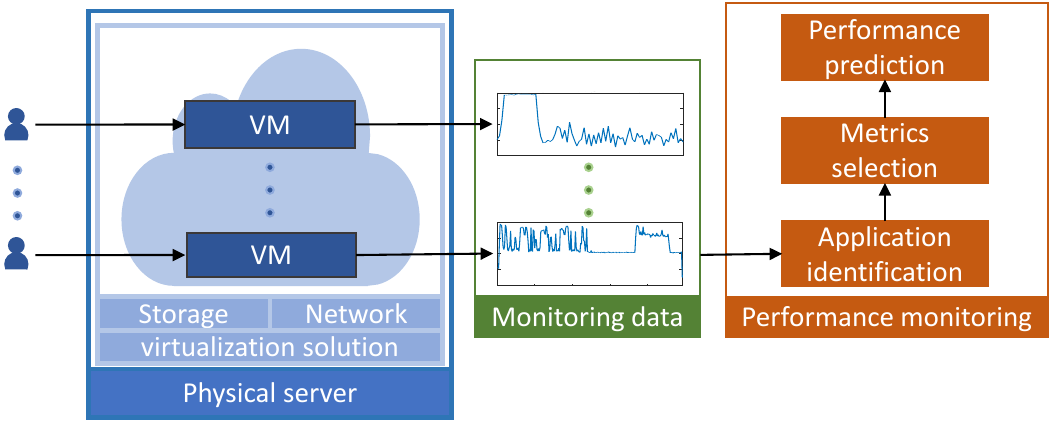}
    \caption{System description and proposed workflow}
    \label{fig:system_description}
\end{figure}

The overall system architecture is presented as the blue box in Figure~\ref{fig:system_description}.
In this work, we focus on a typical configuration of the computing node in public clouds, in which multiple VMs of different types can be executed together, sharing the same computing resources.

\subsection{VM monitoring}
As we discussed previously, state-of-the-art works made simple assumptions on the full knowledge of cloud service clients, i.e., the application running inside the VM is known to the cloud providers~\cite{pham_2017_TCC_executiontimepredict}, or even the source code and specific application running state are known~\cite{lee_2015_compass}. However, such information should be regarded as unknown to cloud providers due to privacy policies, i.e., cloud providers and resource governors have no idea about the application running inside the VM. Thus, lacking knowledge of specification applications running inside the VM and their performance status disables the functionalities of existing VM management schemes~\cite{pham_2017_TCC_executiontimepredict, lee_2015_compass}.

In this paper, we tackle this dilemma by only using the accessible basic hardware usage information of VMs, which can be monitored from the host server without breaching into the VM. These metrics include various aspects of the VM's resource utilization on the host server, such as CPU, cache, memory, hard disk, and network usage information. These metrics are extracted from the host server through performance monitoring tools. Utilizing these hardware metrics does not violate the black-box assumption as all the profiling is done from outside the VM, i.e., on the host server.
Key representatives of these metrics are listed in Table.~\ref{table:libvirt_statistics}.

\begin{table}[t]
\centering
\small
\caption{Representative list of typical collected hardware metrics}
\label{table:libvirt_statistics}
    \begin{tabular}{ c c }
    \toprule
    \textbf{Category} & \textbf{Typical extracted metrics} \\
    \midrule
    \multirow{2}{*}{CPU}  & CPU utilization level (\%) \\
      & Executed instructions (\#)\\
      \midrule
    \multirow{3}{*}{Memory} & LLC misses (\#) \\
     & Available memory space (KB)\\ 
     & Read requests issued for disk usage (\#) \\
      \midrule
    \multirow{2}{*}{Network} & Received packets (Bytes) \\
     & Sent packets (Bytes) \\
   \bottomrule
    \end{tabular}
\end{table}

\subsection{Problem description}

By regarding the VMs as black-box systems, the proposed workflow in this work only takes the hardware statistics data of the VM as the input to predict the application type and its performance status. To achieve the goal, we break the workflow into the following three parts as the orange box illustrated in Fig~\ref{fig:system_description}. As for detail solutions to tackle these problems are introduced in Section~\ref{sec:proposed_method}.

\textbf{Application type identification}:
Previous studies~\cite{pham_2017_TCC_executiontimepredict,lee_2015_compass} assume that the application running inside the VM is known to the cloud providers at runtime, which is unrealistic in public cloud servers. Once removed this invalid assumption, how to predict the performance of the VM remains a question. Therefore, identifying the specific application running inside the VM is the first challenge in public cloud servers and the proposed workflow. In addition, a general VM can run different applications according to the specific needs of users. Therefore, existing management methods failed to adapt to a more dynamic scenario and consider different applications running inside the VM, finally narrowing down their application range. To solve this problem, we propose to regard the VM as a black-box and firstly measure the hardware counter information, including CPU utilization, LLC occupation, memory usage, etc., for VMs on the host physical server. Then, according to the relationship between hardware usage information and the application, we propose to identify the running application inside the VM solely based on the collected hardware counter information.
    
\textbf{Metrics selection}:
After the application is identified, the next step is to predict the application's performance based on measured hardware metrics. However, not all of the sampled metrics are useful for predicting performance. Besides, considering all of the sampled metrics may increase the complexity or even degrade the accuracy of the proposed workflow.
Hence, a proper selection of runtime metrics will lead to a better prediction result.

Due to the application's specific function and its inner logic, different applications demonstrate different utilization levels of the hardware. In particular, memory-intensive applications utilize much more memory resources than computing resources, thus demonstrating much higher memory usage than the CPU utilization level.
From an efficiency perspective, it makes sense to predict the performance level of memory-intensive applications from memory usage metrics instead of non-related metrics such as CPU utilization.

In summary, the first merit of selecting highly-correlated metrics is achieving a less biased prediction method by removing the unrelated metrics. Another benefit of selecting highly-correlated metrics is that we can shrink the prediction model's size and then decrease the complexity of the performance prediction method.
From this perspective, the second step in this work is to perform comprehensive metrics selection analysis for each application.

\textbf{Performance prediction}: Based on the highly-correlated hardware metrics information of the target application, this step first studies the relationship between measured metrics and the application's performance level. Moreover, it enables the framework to predict the performance level from the hardware counter information while considering the interference from other applications and VMs.

The remaining challenge in the performance prediction is to consider workload-induced performance level variation, e.g., a longer execution time of the application may be because of a higher workload level of the application itself without receiving any interference from other VMs. Therefore, workload-induced performance variation should not be regarded as performance degradation.
In this regard, we propose to address this challenge by identifying the application's performance baseline in different workload levels.
Then, a performance degradation level is introduced to normalize the runtime performance level with respect to its performance baseline.
For instance, a performance degradation level near the value one indicates performance degradation is negligible. Otherwise, the application suffers from performance degradation.

Based on the above system and problem description, the proposed machine learning-based performance prediction method to address the existing challenges is introduced in detail in Section~\ref{sec:proposed_method}.

%% file: s5_proposed_method.tex
\section{Machine Learning-based Performance Prediction}
\label{sec:proposed_method}
\begin{figure}[t]
    \centering
    \includegraphics[width=\columnwidth]{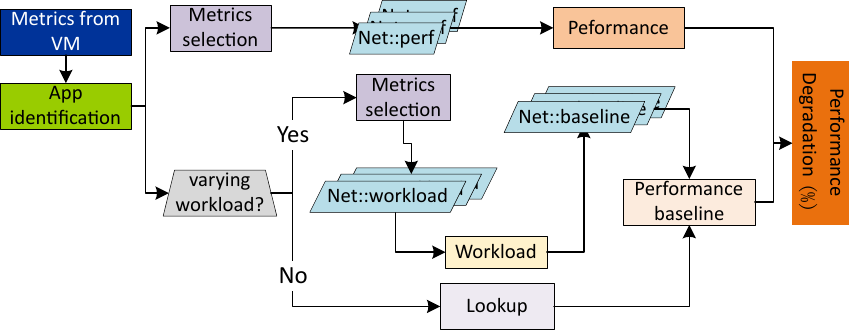}
    \caption{Overall workflow of proposed method}
    \label{fig:workflow_perf_predic}
\end{figure}

We introduce the proposed method in this section by following the workflow illustrated in Fig.~\ref{fig:workflow_perf_predic}. In particular, we introduce a novel application identification method to tackle the dilemma of the black-box nature of cloud VMs in Section~\ref{sec:app_identify}. Then in Section~\ref{sec:metric_selection}, we introduce how to select highly correlated metrics for the purpose of performance prediction. Finally, we present the prediction of performance and performance degradation in Sections~\ref{sec:neuralnetwork} and~\ref{sec:perf_deg}, respectively.

\subsection{Application type identification method}
\label{sec:app_identify}
\begin{figure}[t]
    \centering
    \includegraphics[width=\columnwidth]{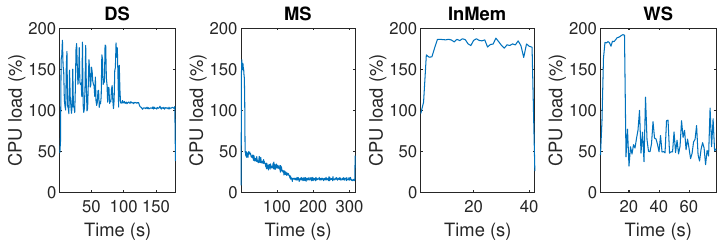}
    \caption{CPU load traces for different CloudSuite benchmarks. When the CPU load is more than 100\%, it implies that more than one core is required to fulfill the performance requirements of the application}
    \label{fig:cpu_traces}
\end{figure}

The proposed application type identification method is based on the fact that different applications have divergent runtime behaviors (e.g., CPU load trace, LLC occupancy). Fig.~\ref{fig:cpu_traces} shows the CPU load traces of four applications from the CloudSuite benchmark~\cite{cloudsuite}, including Data Serving (DS), Media Streaming (MS), In-memory Analytics (InMem), and Web Serving (WS). This selection of cloud benchmarks covers the most representative applications of public clouds~\cite{cloudsuite}, and was used to carry out the experiments. These divergent behaviors indicate clear changes in the specific function and the inner logic of applications.

For clarification purposes, let's consider that only the CPU load metric is used to identify the application. In this scenario, our approach
first builds a ``fingerprint" database ($\mathbb{D}$) with CPU load traces of existing applications.
In the inference step, the unknown application's CPU load trace will be sampled and compared with the traces inside the reference dataset. Finally, the application running inside the VM can be determined based on the similarity analysis.

\begin{figure}[t]
    \centering
    \includegraphics[width=\columnwidth]{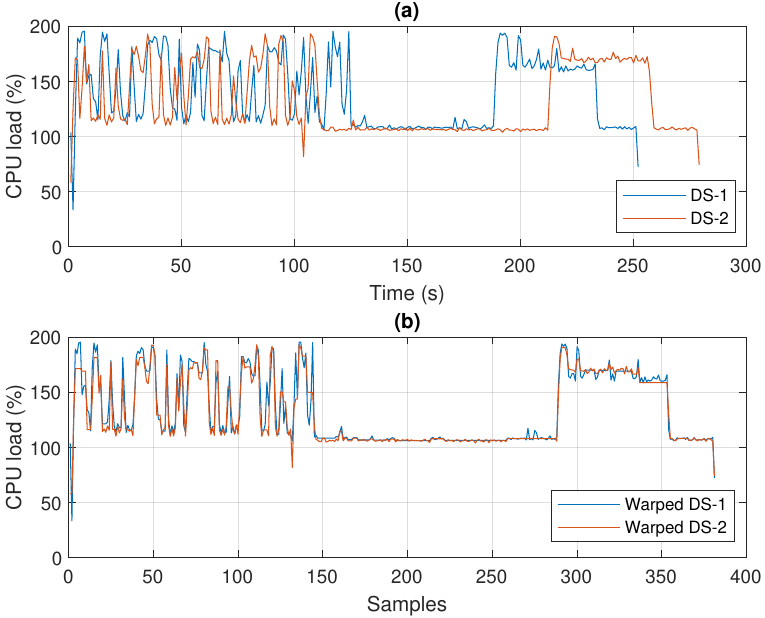}
    \caption{CPU load traces for the application DS (a) Orginal traces (b) DTW processed traces. Note that a load higher than 100\% implies a need for more than one core to respect performance requirements of the application}
    \label{fig:dtw}
\end{figure}

However, the problem of addressing the temporal mismatch between traces remains before any similarity analysis is applied. Indeed, even the traces of the same application can have different shapes because of its own workload variation or the interference the application received from other VMs. Therefore, the same application can give two traces with different dimensions, i.e., $p \in \mathbb{R} ^{1\times M}$ and $q \in \mathbb{R} ^{1\times N}$. For instance, Fig~\ref{fig:dtw}(a) shows two sampled CPU load traces for the same application (DS). The trace DS-2 looks like being stretched when compared with DS-1.

In this work, we propose to use the Dynamic Time Warping (DTW) algorithm~\cite{muller_2007_dtwbook} to eliminate the temporal mismatch between different traces, thus highlighting the maximum similarities between the traces. DTW algorithm uses a cost matrix $C \in  \mathbb{R} ^{M\times N}$ to find the maximum similarity between two variables. The cost matrix is formulated by calculating the Euclidean distance of each pair of elements inside $p$ and $q$, i.e., $C(m, n) = d(p_m, q_n)$. To find the minimum Euclidean distance between two traces, the DTW algorithm finds a minimum cost path from $C(1, 1)$ to $C(M,N)$ such that it has the minimum distance that is the following: 
\begin{equation}
    d_{min} = \sum_{\substack{m \in [1:M] \\ n \in [1:N]}} C(m,n)
\end{equation}
In the process of DTW, input traces will be stretched to eliminate the temporal mismatch and reach the minimum Euclidean distance,
i.e., each element of variables $p$ and $q$ may have a chance to be repeated as many times as necessary to achieve a stretch operation.
After the DTW processing, the original mismatched traces in Fig~\ref{fig:dtw}(a) become warped traces in Fig~\ref{fig:dtw}(b), where the temporal mismatch between two traces is eliminated.
In summary, the DTW removes the temporal mismatch between different traces and highlights the maximum similarity between them, thus making it possible to identify the unknown application accurately.

After highlighting the maximum similarity between traces by the DTM processing, our method uses the Euclidean distance to measure the similarity between two traces. Given two traces after DTW process, $p' = (p'_1, p'_2, \cdots, p'_{M'})$ and $q' = (q'_1, q'_2, \cdots, q'_{M'})$, the Euclidean distance between $p'$ and $q'$ is defined as:
\begin{equation}
\small
    d(p',q') = \sqrt{(p'_1-q'_1)^2 + (p'_2-q'_2)^2 + \cdots + (p'_{M'}-q'_{M'})^2}
\end{equation}
A distance value ($d(p',q')$) of zero indicates a perfect match between the traces $p'$ and $q'$, while the similarity between the two traces are decreasing with the increase of calculated distance value.
Therefore, the unknown trace is identified as the application with the minimum Euclidean distance.
Moreover, we empirically set a distance threshold (i.e., 800) to avoid mistakenly categorizing an unknown application and for the best identification accuracy. If the minimum distance is larger than the set threshold, the application type identification workflow will output ``unknown" because there is no significant similarity between the sampled trace and known applications. And therefore, no performance prediction is made for unknown applications.

\begin{figure}[t]
    \centering
    \includegraphics[width=0.8\columnwidth]{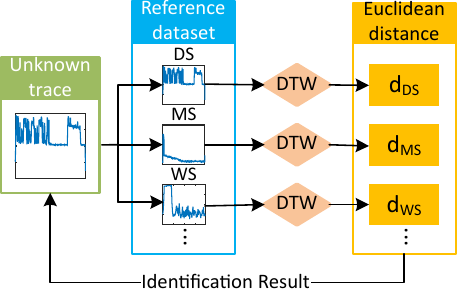}
    \caption{Overall proposed workflow for application type identification}
    \label{fig:workflow_app_identify}
\end{figure}

The overall workflow of application type identification is summarized in Fig.~\ref{fig:workflow_app_identify}. The unknown trace and each trace inside the dataset will be processed through DTW to eliminate the temporal mismatch. Then, the Euclidean distance between the processed signals is calculated. Finally, the unknown trace is identified as the application in the reference dataset with minimum distance.

Please note that hardware metrics for application type identification are not restricted to CPU load traces in this work. Moreover, our proposed workflow enables multiple metrics to be combined through a voting mechanism to improve type identification accuracy. The final application type identification result is the most voted application candidate based on all the hardware metrics monitored from the host server. Once the application has been identified as one of the applications in the dataset, the proposed method can initiate the metrics selection step for the specific application, thus enabling further performance prediction.

\subsection{High-correlated metrics selection}
\label{sec:metric_selection}

\begin{figure}[t]
    \centering
    \includegraphics[width=0.9\columnwidth]{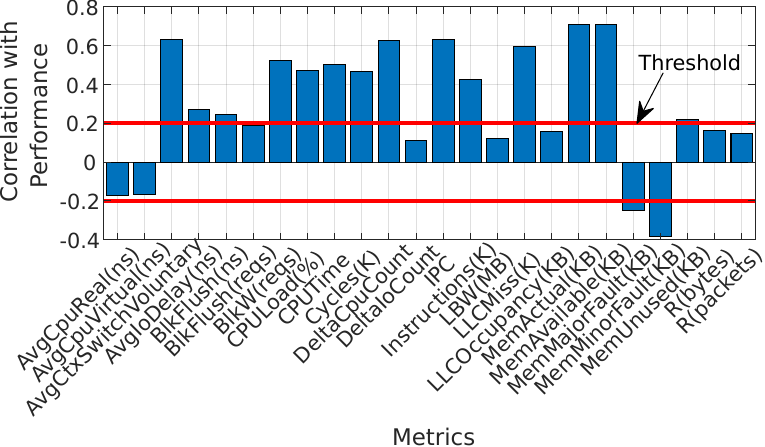}
    \caption{Correlation between hardware metrics and performance for WS}
    \label{fig:correlation}
\end{figure}

The goal of metrics selection is to find hardware metrics that are highly correlated with the target we want to predict. Thus, improving the efficiency and reducing the model complexity of the prediction model. This proposed method is based on correlation analysis of two variables, which are the metric information and the target information (called $a$ and $b$ respectively in the following). We assume both of them have $K$ observations, then the correlation of $a$ and $b$ is defined with the Pearson correlation coefficient~\cite{cox1979theoretical}:
\begin{equation}
\label{eq:pearson_correlation}
\small
    \rho(a,b)=\frac{cov(a,b)}{\sigma_a \sigma_b}= \frac{1}{K}\sum_{i=1}^{K}\frac{a_i-\mu_a}{\sigma_a}\frac{b_i-\mu_b}{\sigma_b}
\end{equation}
where $\sigma_a$, $\mu_a$ and $\sigma_b$, $\mu_b$ are the standard deviation and mean for $a$ and $b$, respectively.
For instance, the correlation between the CPU load level and execution time of the application can be estimated by using the above Eq.~\ref{eq:pearson_correlation} with provided CPU load level information $a$ and execution time information $b$. Similarly, we can enumerate all hardware metrics and performance combinations by substituting the CPU load level metric with other hardware metrics.
As an example, the correlation between metrics and the performance of the application Web Serving (WS) is shown in Fig.~\ref{fig:correlation}. Please note that the correlation has a range of [-1, 1].
The value of exactly -1 or 1 implies a perfect linear relationship between two variables, while values near 0 indicate a weak linear dependency between the variables.
In this work, we select highly correlated metrics that provide an absolute correlation value higher than a set threshold to screen out low-relevant metrics, thus reducing the complicity of the performance prediction workflow.

The proposed metrics selection method works in two aspects as illustrated in Fig.~\ref{fig:workflow_perf_predic}. The only difference between these two usages is the prediction target, i.e., one is for performance prediction for all of the applications and another for workload prediction of applications with varying workloads.

\bhl{Note that none of the hardware metrics exhibit a high correlation (greater than 0.8), as illustrated in Fig.~\ref{fig:correlation}. This indicates the absence of a clear linear relationship between resource usage and performance levels. Therefore, we propose a neural network-based performance prediction method to capture nonlinear patterns. This method will be introduced in the next section.}

\subsection{Neural network-based prediction method}
\label{sec:neuralnetwork}
Artificial neural networks (ANNs), or simply neural networks (NNs), are machine learning models designed to mimic the biological neural networks inside animal brains. A NN is composed of interconnected artificial neurons, which can send information to each other according to the input it receives and the connections among the neural network. A NN usually needs to be trained from a large set of given inputs and outputs. Then the trained network can be applied to estimate the output from the given input.

The NNs used in this work are all well trained before being plugged into the proposed workflow. The training phase aims to accurately predict output $y$ from the given input information $x$. In this work, we utilize the widely used Levenberg-Marquardt backpropagation algorithm to train NNs~\cite{wilamowski2010lm}. Backpropagation optimizes the loss function with respect to the weights parameter used in the network. The weights of the NN will be set randomly before the training. During the training phase, the weights will be updated by the backpropagation algorithm to minimize the lost function. Then, for each training iteration, the training sample is composed of an input set $X$ and output set $Y$. Based on the input $X$, NN gives its own prediction $\hat{Y}$ according to its structure and weights parameter. During the training phase, the backpropagation algorithm adjusts the weight parameter of the NN to approximate $\hat{Y}$ to $Y$. Note that our approach only uses internal performance and workload metrics of the application in the initial training phase. For the inference phase, the NN will provide the prediction by only using low-level hardware metrics. Therefore, our method is agnostic to each application's internal metrics and follows the required black-box model setting for VMs performance inference.

The input and output of the NN, i.e., $X$ and $Y$, differ from the specific application or purpose. In this work, three different purposes of NNs are used in the workflow described in Fig.~\ref{fig:workflow_perf_predic}, where:
\begin{itemize}
    \item \textbf{Net::performance}: The output $Y$ of the NN is a scalar value to indicate the performance level of the application. It differs from application to application. For instance, the performance metric for the application Data Serving (DS) is the execution time, while it is operations per second for the application Web Serving (WS). The input $X$ contains metrics that show high correlations with the performance metric, and it is given by the metric selection analysis from Section~\ref{sec:metric_selection}.
    \item \textbf{Net::workload}: The output $Y$ of the NN is a scalar value that indicates the workload level of the application. It also depends on the specific application. For instance, the DS uses the operation count as its workload indicator, while WS's workload is measured by the user number. Similar to \textbf{Net::performance}, the input $X$ contains metrics that show high correlations with the workload level of the application, and it is given by the prior metrics selection step.
    \item \textbf{Net::baseline}: The input of this performance baseline prediction NN is the output \textbf{Net::workload}, which is the workload level of the application, while the output of \textbf{Net::baseline} is the performance baseline of the application under the specific workload level. In this work, we regard the best performance the application can achieve on the target system in the specific workload level as the performance baseline.
\end{itemize}

\bhl{In this work, we did not choose deep neural networks (DNN) for the following reasons. First, deeper networks actually possess greater learning capacity, enabling them to effectively capture complex patterns within the data. However, it is important to acknowledge that deeper networks are also more susceptible to overfitting training data, resulting in a lack of generalization when applied to the test set. Second, the neural networks trained in this work go through hyperparameter optimization on the number of layers (exploration space up to hundreds) for the best prediction results. Therefore, irrespective of the NN vs. DNN terminology, we have optimized the parameters of our neural networks to maximize prediction performance while avoiding the larger computing overhead brought about by more complicated deep neural networks.}

In this work, each application has an individual set of the above NNs. The purpose of this configuration is to have a robust and elastic prediction workflow. Otherwise, the proposed workflow will have numerous drawbacks if all applications use the same all-purpose NN set.
First, the proposed workflow will suffer from a high training overhead because the trained neural network needs to be retrained from scratch to support a new application.
Second, the retrained NN cannot guarantee robustness or remain consistent for existing applications. i.e., the newly trained network's performance cannot be guaranteed the same as the previously trained network.
Last but not least, trained DNN needs to be redesigned to deal with increasing complexity while supporting both existing and new application types.
In summary, it does not make sense to train an all-purpose NN set for all applications. Instead, we choose to train a dedicated set of NN for each application to make the framework more elastic and robust in supporting possibly new applications while reducing the training complexity and overhead.

\subsection{Performance degradation prediction}
\label{sec:perf_deg}
As introduced before, the predicted performance level of the application cannot fully reflect the interference VM received due to the existence of workload-induced performance variation. To address this challenge, we propose an NN-based prediction method for performance degradation. First, the proposed method will check if the application can have variable workload levels after the running application is identified. Suppose the application belongs to one without variable workloads, which indicates the application has a constant performance baseline. Therefore, the proposed method can directly give the performance baseline from the knowledge of these applications. Please note that the performance baseline is the best performance the application can achieve in the system without interference.

Otherwise, if the application has variable workloads, i.e., DS and WS, the proposed method will first select the metrics which are highly correlated with the workload level of the application. By using these chosen metrics, the workload level is predicted using the NN \textbf{Net::workload}. Then, another NN, \textbf{Net::baseline} introduced in Section~\ref{sec:neuralnetwork}, is used to predict the performance baseline from the predicted workload level. Finally, the performance degradation is calculated by dividing the predicted performance level by the performance baseline in the current workload level. If the application is running without interference, the performance degradation always gives a value of one despite the workload level varying, thus indicating the runtime performance is exactly the same as the performance baseline. Otherwise, the performance degradation will be larger than one to indicate the runtime performance is inferior to the performance baseline because of the interference. Finally, the resource governor can utilize the predicted performance degradation value to schedule VMs on the server for better performance and lower interference.

\begin{algorithm}[t]
\SetAlgoLined
\SetKwInput{Input}{Input}
\SetKwInput{Output}{Output}
\Input{System runtime metrics: $Metrics$}
\Output{Performance degradation level: $Perf_{deg}$}

Identify the application from the runtime metrics\;

$Metrics_{perf} \leftarrow$ metrics highly correlated with the runtime performance level\;

$Perf \leftarrow Net::performance(Metrics_{perf})$ \;

\eIf{The application has variable workloads}{
   $Metrics_{wkload} \leftarrow$ metrics highly correlated with the runtime workload level\;
   $Workload \leftarrow Net::workload(Metrics_{wkload})$ \;
   $Perf_{base} \leftarrow Net::baseline(Workload)$ \;
   }{
   $Perf_{base} \leftarrow$ best performance the application can achieve
in the system \;
   }

$Perf_{deg} = \frac{Perf}{Perf_{base}}$\;

 \caption{Neural  Network-based  Performance  Degradation Prediction}
 \label{alg:workflow}
\end{algorithm}

In summary, the overall algorithm of the proposed method is described in Algorithm~\ref{alg:workflow}. The algorithm takes measured metrics from the host server as the input and predicts the performance degradation of the application.
In the first step, the proposed application type identification method is used to identify the running application (Line 1).
Then, only the metrics that are highly correlated with the runtime performance level of the application are selected by the proposed method (Line 2).
Given the selected runtime metrics, the runtime performance level of the application is predicted by using the trained NN \textbf{Net::performance} (Line 3).
As for performance baseline prediction, the proposed method will first check if the application has variable workloads.
If yes, metrics highly correlated with the workload level will be selected first (Line 5), and then the runtime workload level will be predicted by using the NN \textbf{Net::workload} (Line 6). After the workload level is predicted, the corresponding performance baseline under this workload is predicted using the NN \textbf{Net:baseline} (Line 7).
Otherwise, if the application has a constant workload, the performance baseline is the best performance the application can achieve in the system without interference. In this circumstance, we directly get the performance baseline from the knowledge base (Line 9).
Finally, after the performance and performance baseline are both predicted, the performance degradation is calculated by dividing performance by the performance baseline (Line 11).

%% file: s6_exp_setup.tex
\section{Experimental setup}
\label{sec:exp_setup}

\subsection{Server and VM configurations}

The cloud infrastructure used in this work is composed of two different types of servers, listed in Table~\ref{table:server}. 
The first server, S1, corresponds to a computing node provided by the Huawei public cloud infrastructure located in Xi'an, China.
With the purpose of verifying the versatility of the proposed method, another server, S2, is configured in our computing experimental facility in the datacenter of EPFL, by following the configuration listed in Table~\ref{table:server}.

In this work, the virtualization solution is based on OpenStack, which has been vastly deployed for cloud services~\cite{Openstack}. OpenStack enables cloud service providers to create and manage multiple VMs on a single computing server thanks to its key components, including Ceph for distributed block storage~\cite{Ceph}, Open vSwitch for networking~\cite{OpenvSwitch}, and KVM for launching virtual machine~\cite{KVM}.

A total of three different types of VM are used in this work, which are listed in Table~\ref{table:vm}. The three tiers of VM configurations correspond to VM configurations widely employed by popular cloud service providers among Amazon Elastic Compute Cloud, Microsoft Azure, Google Compute Engine, and Huawei Elastic Cloud Server. Then, to provide a fair comparison among different VMs and servers, the same operating system, Ubuntu 20.04, is installed on all VMs.

\bhl{With contemporary virtualization solutions and monitoring tools, such as libvirt and Linux perf-kvm, comprehensive support is available for monitoring individual hardware resource usage of each guest VM atop the whole physical server. Hence, in this work, we employed the existing libvirt monitoring tools~\cite{libvirt} to account for the hardware resource usage for a single VM on the entire physical server while serving multiple VMs, allowing us to scrutinize the hardware resource utilization of each VM individually and export the corresponding data into a separate database for each VM.}

\bhl{In addition, the monitoring tools are employed on the host server (outside the VMs) to monitor each colocated VM running on the same physical server. This choice not only aligns with the black-box assumption adopted in our study but also allows for resource usage accounting and exclusive partitioning outside the VMs. Consequently, eliminating any potential errors associated with in-guest VM measurements.}

\begin{table}[]
\centering
\caption{Server configurations used in the experimental section}
\label{table:server}
\begin{tabular}{c c c c}
\toprule
Server & CPU    & Memory    & OS    \\ \midrule
S1       & \makecell{ 2x Intel Xeon \\E5-2620 v3 @ 2.4GHz}    & 128GB ECC & EulerOS 2.0 \\ 
S2       & \makecell{2x Intel Xeon \\Gold 6242R @ 3.1GHz}     & 384GB ECC & CentOS 7.9 \\ 
\bottomrule
\end{tabular}
\end{table}

\begin{table}[]
\centering
\caption{VM configurations tested in the experiments}
\label{table:vm}
\begin{tabular}{c c c c}
\toprule
VM name & CPU    & Memory & OS    \\ \midrule
VM1      & 2x vCPU    & 4GB & Ubuntu 20.04 \\ 
VM2       & 4x vCPU    & 4GB & Ubuntu 20.04 \\ 
VM3       & 4x vCPU    & 8GB & Ubuntu 20.04 \\ \bottomrule
\end{tabular}
\end{table}

\subsection{Comparison methods}
In this work, we compare the proposed method with three state-of-the-art prediction methods for cloud applications.

\subsubsection{Cao \textit{et al.}~\cite{cao_2018_dbworkload}}

The first comparison method is from Cao \textit{et al.}~\cite{cao_2018_dbworkload}. It can predict the incoming workload for servers based on a decision tree method. The decision tree is a tree-like model of decisions and their possible consequences, where each non-leaf node represents the condition of the characteristic attributes and each leaf node indicates a category. The decision-making process starts from the root node, then makes a decision and chooses a path until it reaches a leaf node and gives the final decision and prediction. The prediction method proposed by Cao \textit{et al.}~\cite{cao_2018_dbworkload} can provide predicted incoming workload information to the resource governor, which then handles server configurations to tackle the possible workload stress. 
In this work, we adapt the workload prediction method proposed by Cao \textit{et al.}~\cite{cao_2018_dbworkload} to consider performance prediction for the VM in the experiment. 
Besides, we use the Bayesian optimization method to optimize the minimal leaf size of this method, thus achieving its best performance.

\subsubsection{Pham~\textit{et al.}~\cite{pham_2017_TCC_executiontimepredict}}
Pham~\textit{et al.}~\cite{pham_2017_TCC_executiontimepredict} propose to use a random forest method to predict task execution time in the cloud. Random forest packs multiple decision trees to make the decision by utilizing a voting mechanism of all the individual decision tree models. Thus, it addresses the overfitting problem suffered by the solution proposed by Cao \textit{et al.}~\cite{cao_2018_dbworkload}
However, this work does not consider the interference-induced performance variation of the application. In other words, the execution time solely depends on the server type and VM configuration. Therefore, it is unfair to compare their proposed random forest method because it can only give a fixed performance value once the cloud platform, VM type, and application are determined without respecting the interference the application received. In the end, to achieve a fair comparison, we adapt the prediction method proposed by Pham~\textit{et al.}~\cite{pham_2017_TCC_executiontimepredict} in predicting the runtime performance level of the application by using the application's runtime metrics. Moreover, to achieve the best performance for this method, we use the Bayesian optimization method to optimize its hyperparameters (i.e., number of trees and minimal leaf size).

\bhl{
\subsubsection{Bader~\textit{et al.}~\cite{bader2022lotaru}}
Bader \textit{et al.}~\cite{bader2022lotaru} introduced an innovative technique called Lotaru, aims to estimate task runtimes in scientific workflows deployed on heterogeneous cloud servers. Lotaru operates in several steps: initially, it performs comprehensive profiling of all servers using a set of benchmarks. Subsequently, it takes advantage of the obtained measurements to train a Bayesian regression model. This model enables the prediction of a task's runtime based on its input size. Moreover, Lotaru fine-tunes the predicted runtimes for each different server in the cluster, incorporating the benchmarking results as specific adjustments.
To facilitate a detailed analysis, we implemented the Lotaru method that uses workload as input for estimating task performance, named as \textit{Lotaru-workload}. Additionally, in order to ensure a fair comparison, we extend the Lotaru to support runtime hardware metrics, the same ability as other comparison methods, for estimating task performance, named as \mbox{\textit{Lotaru-trace}}.
}

\subsection{Cloud benchmarks}

\begin{table*}[hbt!]
\centering
\caption{\bhl{Applications used in the experiments}}
\cb
\begin{tabular}{c c c c}
\toprule
Application name & Purpose  & Performance metric    & Performance baseline \\ \midrule
Data Serving (DS)      & stress the data store and serving server & execution time (s) & [40.2 100.0]    \\ 
Web Serving (WS)       & stress the throughput and latency of web services & operations/s  & [3.0 8.6]   \\ 
Media Streaming (MS)       & stress the server with video streaming applications  & requests/s  & 25.7   \\ 
In-Memory Analytics (InMem)      & stress the server with the recommender algorithm    & execution time (s) & 35.8 \\ 
Redis benchmark       & evaluate the performance of the data structure server    & requests/s & 5.4e4  \\
\bottomrule
\end{tabular}
\label{table:benchmarks}
\end{table*}

Target cloud benchmarks are selected from CloudSuite~\cite{cloudsuite}, an open-source benchmark suite of emerging scale-out workloads. It contains a collection of representative application categories commonly found in today's data centers. We also added the Redis benchmark~\cite{redis_bench} to the benchmark set to test the performance of the data structure server and also enlarge the variability of target cloud applications. In summary, target cloud applications used in this work are listed in Table~\ref{table:benchmarks}. \bhl{Note that both DS and WS applications exhibit a performance baseline range due to their variable workload levels. On the contrary, the remaining three applications have a fixed performance baseline as a result of their fixed workload level.}

Different from the non-interference setting used in the comparison method~\cite{pham_2017_TCC_executiontimepredict} where the application runs without receiving interference from other activities of the server, the interference between applications and VMs is well considered in our work. To incorporate as many interference combinations as possible, a dynamic scenario is considered in this work. During the experiment, there are five VMs running concurrently inside the server.
This scenario definition guarantees that each VM receives a moderate stress level, as we showed previously in our motivation example in Fig.~\ref{fig:ds_perf_dist}. In this figure, the performance of DS can degrade more than eight times.
If more VMs are running on the server, the VMs always experience some performance degradation. On the contrary, a fewer number of VMs will not provoke enough interference to each other.
Each VM has a total of six operation modes (0-5), as listed in Table~\ref{table:operation_mode}. The operation mode for each VM is randomly chosen at runtime. If the mode is 0, the VM is idle and does not run any application for three minutes. Otherwise, the VM runs the application listed in Table~\ref{table:operation_mode}.
With five VMs concurrently running on the server and following the six operation modes setting, the designed experiment can cover every running application combination and interference level for each VM.

In this work, we have collected over 10,000 instances of applications running on servers. The neural network in the proposed method takes 70\% of data samples for training, and each validation and test phase takes 15\% of overall data samples.
For a fair comparison, both comparison methods are also tested with the same 15\% samples. Considering comparison methods, based on decision tree, random forest, and bayesian regression techniques, do not have the validation phase as the neural network. They are all trained with 85\% data samples (i.e. 15\% more training samples than our proposed method).

\begin{table}[]
\centering
\caption{VM's operation modes and relevant actions}
\begin{tabular}{c c}
\toprule
Mode & Action  \\ \midrule
0      & Idle for 3 minutes    \\ 
1       & Run DS    \\ 
2       & Run WS    \\ 
3      & Run MS \\ 
4      & Run InMem \\ 
5       & Run Redis \\ \bottomrule
\end{tabular}
\label{table:operation_mode}
\end{table}

%% file: s7_exp_results.tex
\section{Experimental Results}
\label{sec:exp_results}

In this section, we first target the server S1 and VM1 to evaluate the accuracy of the application type identification method proposed in Section~\ref{subsec:app_identify}. 
Then, the performance prediction results are compared against different methods in Section~\ref{subsec:perf_predict}. 
Section~\ref{subsec:perfdeg_predict} investigates the prediction of performance degradation by using the proposed method. To verify the proposed method can also fit into various VM and Server configurations, the workflow versatility is examined in Section~\ref{subsec:versatility} with new server S2 and VM types 2 and 3. In addition, the sampling and computation overheads of the proposed method are studied in Sections~\ref{subsec:accuracy_sampling} and~\ref{subsec:overhead} respectively.

\subsection{Application type identification}
\label{subsec:app_identify}
The proposed application type identification method compares the trace of the unknown application with the reference dataset to find its identity. Therefore, a comprehensive reference dataset that contains many runtime features of the application has a crucial impact on the accuracy of the identification process. The relationship between the complexity of the reference dataset and identification accuracy is studied in this section. As an automatic method without expert knowledge, the complexity of the reference dataset is expressed in the number of traces contained in the dataset. With a larger number of traces in the reference dataset for each application, the application's possible behaviors are more likely to be captured by the reference dataset and identified with the proposed method.
\begin{figure}[t]
    \centering
    \includegraphics[width=\columnwidth]{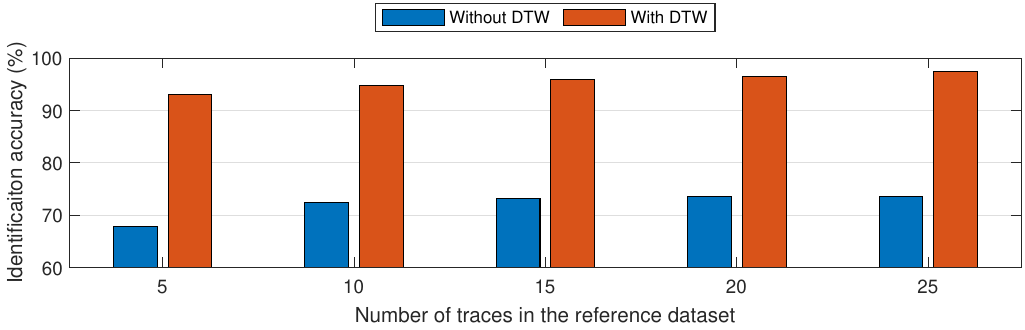}
    \caption{Validation results of our proposed application type identification algorithm}
    \label{fig:with_out_dtw}
\end{figure}

Fig.~\ref{fig:with_out_dtw} collects validation results of the proposed application type identification method (with DTW) along with the identification method without using DTW. The evaluation starts with only one trace for each application, i.e., five traces in the reference dataset for all of the five applications used in this work. Thanks to our proposed use of DTW technology, we achieve very high accuracy, 93\%, in the application type identification process even with only one trace for each application. However, the identification method without using DTW can only achieve 68\% accuracy in the same setting. With increasing the number of traces to 20, i.e., four traces for each application, the identification results increase to 97\% with the proposed method. Beyond 20 traces, the accuracy stops growing with the number of traces in the reference dataset. In comparison, the identification method without DTW reaches a plateau only at a 74\% accuracy when increasing the number of traces. In summary, Fig.~\ref{fig:with_out_dtw} demonstrates that the proposed DTW-based application type identification method can achieve less than 3\% error in identifying running applications.

\bhl{Hardware configurations can differ across various physical servers, thus influencing the resource usage level of applications across different servers. Therefore, the proposed approach requires a retraining step to adapt to these variations and ensure high accuracy. For the new server configuration, it is necessary to re-sample and store the application traces within a new reference dataset target. Note that this retraining process, only needs to be done when server configuration changes. Once retrained, it can be employed on the same type of server without additional training overhead. In this work, we verified the proposed application type identification method for both types of server S1 and S2, all achieving a precision greater than 97\% despite variations in server configurations.}

\subsection{Performance prediction with interference}
\label{subsec:perf_predict}

In this section, we compare the performance prediction results between the proposed method and comparison methods.
To make a fair comparison between different methods, the proposed method is compared with the state-of-the-arts only in performance prediction. In other words, only the neural network \textbf{Net::performance} in the proposed method is used in the following experiments.

\begin{figure*}[!t]
\centering
\begin{minipage}[t]{0.23\linewidth}
  \centering
  \includegraphics[width=\columnwidth]{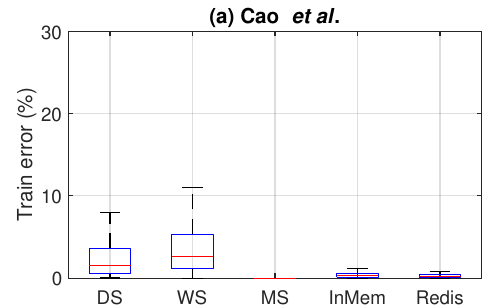}

\vspace{2mm}
\end{minipage}
\quad
\begin{minipage}[t]{0.23\linewidth}
\centering
\includegraphics[width=\columnwidth]{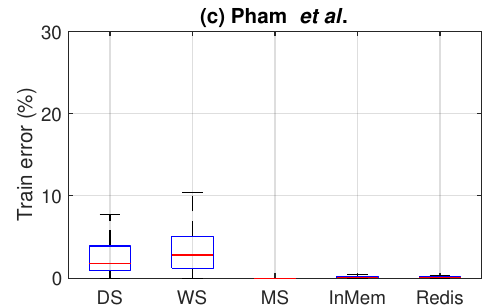}
\end{minipage}
\quad
\begin{minipage}[t]{0.23\linewidth}
\centering
\includegraphics[width=\columnwidth]{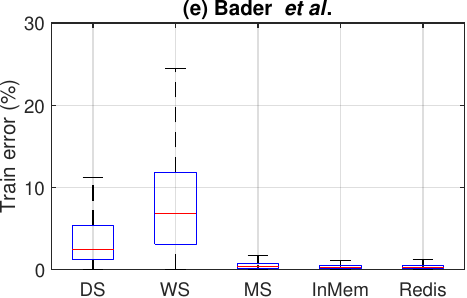}
\end{minipage}
\quad
\begin{minipage}[t]{0.23\linewidth}
\centering
\includegraphics[width=\columnwidth]{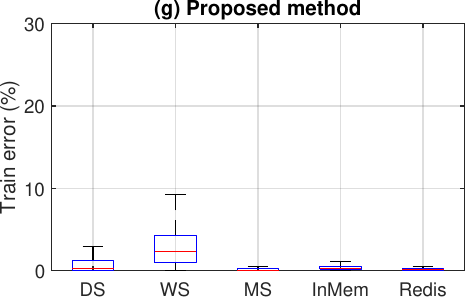}
\end{minipage}
\quad
\begin{minipage}[t]{0.23\linewidth}
  \centering
  \includegraphics[width=\columnwidth]{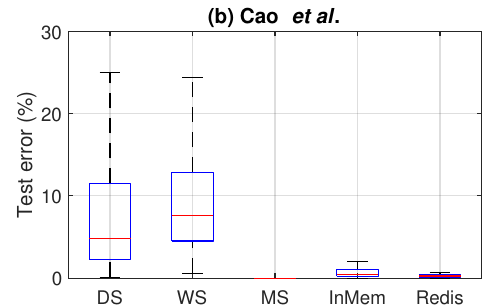}
\end{minipage}
\quad
\begin{minipage}[t]{0.23\linewidth}
\centering
\includegraphics[width=\columnwidth]{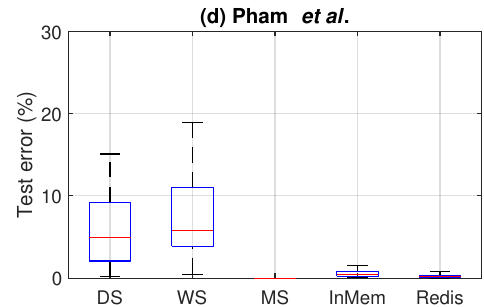}
\end{minipage}
\quad
\begin{minipage}[t]{0.23\linewidth}
\centering
\includegraphics[width=\columnwidth]{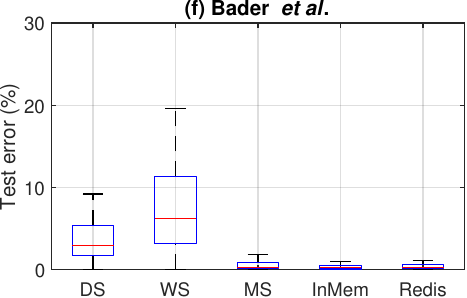}
\end{minipage}
\quad
\begin{minipage}[t]{0.23\linewidth}
\centering
\includegraphics[width=\columnwidth]{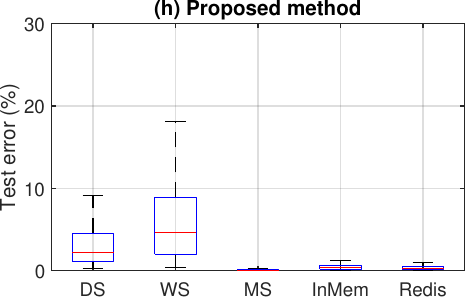}
\end{minipage}
\quad
\caption{\bhl{Comparison of different methods for performance prediction. Note that DS and WS present a varying workload, conversely to other benchmarks}}
\label{fig:performance_predic_boxplot}
\end{figure*}

A set of boxplot figures listed in Fig.~\ref{fig:performance_predic_boxplot} are used to show an overall graphical comparison of different methods in terms of training and test performance. The first row of Fig.~\ref{fig:performance_predic_boxplot} shows the training phase error for different methods, while the second row focuses on the test error. Inside each boxplot, there are results for the five applications. The blue box of the quartile captures the error data in the range of 25\% to 75\%. The red line inside the blue box is the median of the error. Therefore, a lower and smaller blue box indicates a lower prediction error and a smaller standard deviation.
\bhl{Note that in the figure, the method presented by Bader \textit{et al.} primarily focuses on the \textit{Lotaru-trace} approach. In our study, we also implemented the \textit{Lotaru-workload} method. However, we observed that the train and test errors for \textit{Lotaru-workload} method are significantly higher, reaching above 39\% and 20\% for DS and WS, respectively.
This can be attributed to the fact that the application's performance is not solely determined by the workload itself, but also by the interference the application experiences. Modeling the workload alone fails to capture the complete performance characteristics of the application. Consequently, by incorporating runtime hardware metrics using the \mbox{\textit{Lotaru-trace}} approach, it can achieve more accurate prediction results at a similar level compared to other methods.}

In summary, all four methods have similar performance and high accuracy in predicting performance for applications without variable workloads (i.e., MS, InMem, and Redis). However, the comparison methods fall short in performance prediction for applications with variable workloads (i.e., DS and WS). In contrast, the proposed neural network method demonstrates its advantage and shows the best accuracy in predicting performance for application DS and WS.

\subsection{Prediction of the performance degradation}
\label{subsec:perfdeg_predict}

In the previous section, the advantages of the proposed method in performance prediction are well demonstrated. However, only the application's performance level cannot fully reflect the interference the VM received. For instance, the performance level of application WS also varies with the number of users, i.e., workload level. Therefore, a small number of users or a VM receiving interference from other VMs, can both lead to a low performance level of WS. Therefore, we proposed a novel method to tackle the existing problem. Since the comparison methods do not support the prediction of performance degradation, this section mainly focuses on verifying the proposed method in the prediction of performance degradation.

The prediction results are collected in Table~\ref{table:perf_deg_nn}.
The proposed method achieves less than 5\% error in the training phase and 7\% error in the test phase for both DS and WS. As for the applications without variable workloads (i.e., MS, InMem, and Redis), the proposed method predicts performance degradation with less than 0.5\% and 1.2\% error in the training and test phases, respectively.
The above results fully verify that the proposed method outperforms the state-of-the-art methods in the application range and can predict the performance degradation for the VM with high accuracy.

\begin{table}[]
\renewcommand{\arraystretch}{1.2}
\centering
\caption{Accuracy of the proposed method in the prediction of performance degradation}
\label{table:perf_deg_nn}
\begin{tabular}{c c c c c c c }
\toprule
 \multirow{2}{*}{App} & \multicolumn{3}{c}{Train error (\%)}  & \multicolumn{3}{c}{Test error (\%)}   \\ 
\cmidrule(lr){2-4} \cmidrule(lr){5-7}
 & mean & max & std  & mean & max & std    \\ \midrule
 
DS      & 4.5   & 44.7  & 5.6   & 5.4  & 26.0  & 5.2    \\ 
WS      & 4.6   & 27.0  & 4.3   & 6.8  & 32.1  & 6.1     \\ 
MS      & 0.5   & 7.6  & 0.9   & 1.2   & 16.4  & 2.9    \\ 
InMem   & 0.4   & 3.3  & 0.4   & 0.6   & 7.4  & 0.8 \\ 
Redis   & 0.2   & 1.3  & 0.2   & 0.3   & 1.7  & 0.3 \\
\bottomrule
\end{tabular}
\end{table}

\subsection{Workflow versatility: porting the solution to different servers and VM types}
\label{subsec:versatility}
In this section, we configure another new server to verify that the proposed workflow can be adapted to different servers. Besides, we also configure different types of VMs running on the new server to further prove that the proposed method can also work with different types of VMs.

\subsubsection{New server}

\begin{figure*}[!t]

\centering
\begin{minipage}[t]{0.31\linewidth}
  \centering
  \includegraphics[width=\columnwidth]{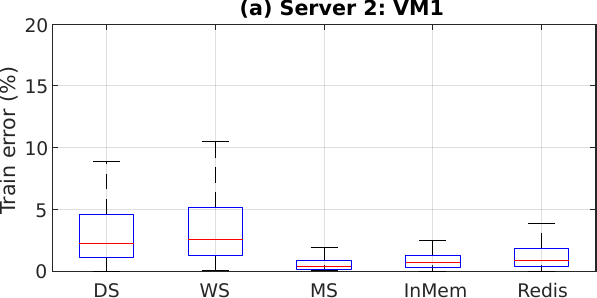}

\vspace{2mm}
\end{minipage}
\quad
\begin{minipage}[t]{0.31\linewidth}
\centering
\includegraphics[width=\columnwidth]{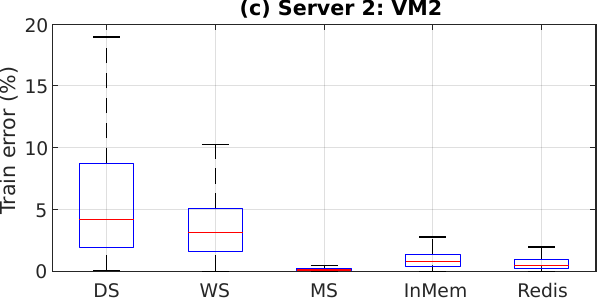}
\end{minipage}
\quad
\begin{minipage}[t]{0.31\linewidth}
\centering
\includegraphics[width=\columnwidth]{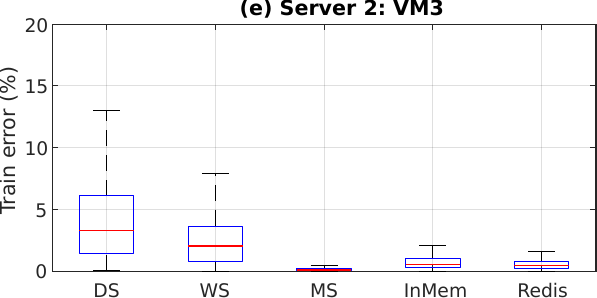}
\end{minipage}
\quad

\begin{minipage}[t]{0.31\linewidth}
  \centering
  \includegraphics[width=\columnwidth]{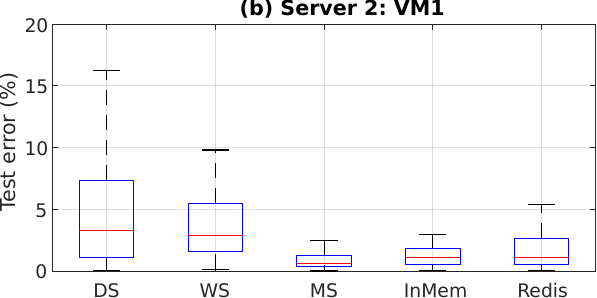}
\end{minipage}
\quad
\begin{minipage}[t]{0.31\linewidth}
\centering
\includegraphics[width=\columnwidth]{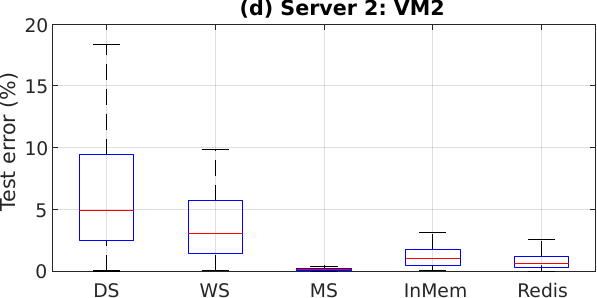}
\end{minipage}
\quad
\begin{minipage}[t]{0.31\linewidth}
\centering
\includegraphics[width=\columnwidth]{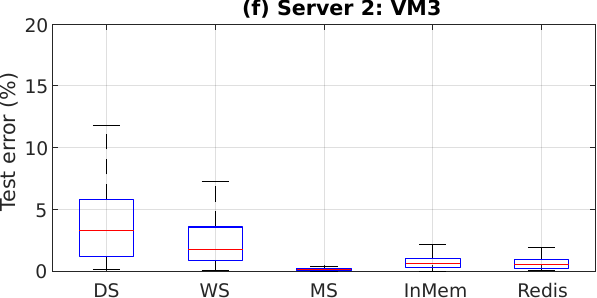}
\end{minipage}
\quad
\caption{Prediction error of the proposed method tested under different server and VM configurations}
\label{fig:perfd_s2}
\end{figure*}

The configuration of the new server (i.e., Server 2), is listed in Table~\ref{table:server}. Other than the server hardware configuration, we keep all other settings the same as the previous experiments.
The new server, S2, has much better performance than another server, S1, because it is equipped with a faster CPU, more memory, etc. For instance, our experiments demonstrate that the Redis benchmark achieves a 62\% average performance improvement on server S2 than S1.
After collecting the experimental samples, the training and test errors are illustrated in the boxplot Figs.~\ref{fig:perfd_s2} (a) and (b). The average prediction error is below 7\% for all of the applications in both training and test phases.
Therefore, a similar error with the previous results for Server configuration S1, shown in Table~\ref{table:perf_deg_nn} of Section~\ref{subsec:perfdeg_predict}, demonstrates that the proposed method works well for different server configurations.

\subsubsection{New VMs}

In addition to the new server configuration, the proposed workflow was also tested with different VM configurations in this work. Different types of VMs, as introduced in Section~\ref{sec:exp_setup}, are configured and then sampled on Server S2.
Based on the experiments with these new VM configurations, Figs.~\ref{fig:perfd_s2} (c) to (f) illustrate the prediction error for both VM2 and VM3. The prediction error distributes in a similar range for all of the three different VM types, i.e., lower than 5\% in terms of median prediction error. This observation demonstrates that the proposed workflow can be extended to new server configurations and different VM types widely used in public clouds without compromising prediction accuracy.

\subsection{Trade-off between sampling time and prediction accuracy}
\label{subsec:accuracy_sampling}
\begin{figure}[t]
    \centering
    \includegraphics[width=0.9\columnwidth]{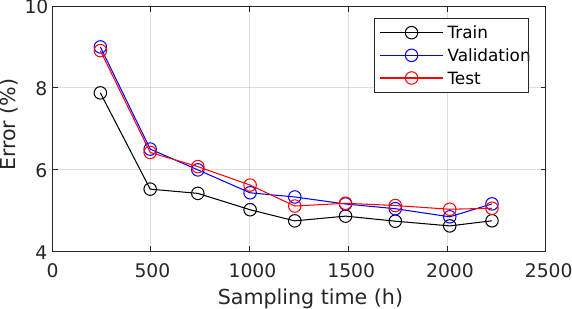}
    \caption{Trade-off between sampling time and prediction accuracy}
    \label{fig:tradeoff_samplingtime_accuracy}
\end{figure}

Machine learning methods require adequate samples of the target application to achieve good prediction accuracy. In this section, we regard DS as a case study to investigate the trade-off between the sampling time and prediction accuracy, as illustrated in Fig.~\ref{fig:tradeoff_samplingtime_accuracy}. In the beginning, the prediction error, in terms of training, validation, and test errors, sharply decrease thanks to the increase in sampling time. This phenomenon indicates that the neural network can achieve better prediction accuracy with more samples for training purposes. However, the prediction accuracy improves limitedly after the sampling time over 1000 hours, when the neural network receives an adequate amount of data samples for the training purpose.

Please note that the sampling time of 1000 hours does not necessarily require the server to run for 1000 hours. It can be achieved with 5 VMs concurrently running on the server for 200 hours. Thus, making it feasible to get adequate application samples in a short period of time with multiple VMs.

\subsection{Computation overhead analysis}
\label{subsec:overhead}
\bhl{Note that the training time belongs to the offline time overhead. Once trained, the model can be deployed on multiple servers for execution without incurring additional training time. Therefore, offline training time is usually neglected in related work and in this work. Moreover, training time is much smaller than any of the recent (large) AI/ML models (e.g., Transformers, etc.), so the overhead for this training process can be considered very limited.}

The runtime computational overhead of different methods for performance prediction is studied and listed in Table~\ref{table:overhead}. Based on the results shown in the table, the proposed neural network-based performance prediction method shows the largest computation overhead due to its higher complexity and accuracy. In comparison, \bhl{both the methods proposed by Bader \textit{et al.} and Pham \textit{et al.} can achieve less computation overhead because of the non-complicated underlying prediction techniques, which are Bayesian regression method and Random tree, respectively. The method proposed by Cao \textit{et al.} has the minimum computation overhead because of the Decision tree it utilized.} However, the overhead of the neural network is still in the \SI{}{\us} scale, being negligible compared to the application's execution time, which usually ranges from minutes to hours. \bhl{Therefore, the runtime inference time of the proposed method is also not a concern. Moreover, with this negligible computing overhead, our proposed method brings the ability to discern different application types running in a black-box scenario of public servers and predict applications' performance degradation level with the awareness of workload variation.}

\begin{table*}[h]
\renewcommand{\arraystretch}{1.2}
\centering
\caption{Computation overhead of different methods in \SI{}{\us}}
\label{table:overhead}
\begin{tabular}{c c c c c c c c }
\toprule
 &  DS &  WS &  MS & InMem & Redis & DS Deg & WS Deg \\
 \cmidrule(lr){2-6} \cmidrule(lr){7-8}

This work      & 4.8   & 5.2  & 13.0   & 15.6  & 13.0  & 16.7  & 16.1\\

\bhl{Bader~\textit{et al.}}   & \cb 4.7  &\cb 4.5 &\cb 4.0 &\cb 4.1 &\cb 3.6  &\cb -  &\cb - \\

Pham~\textit{et al.}      & 2.8   & 3.1  & 5.1   & 7.1  & 6.3  & -  & - \\

Cao \textit{et al.}   & 0.4   & 0.3   & 0.7   & 1.2   & 1.1   & -   & - \\

\bottomrule
\end{tabular}
\end{table*}

%% file: bio.tex
\begin{IEEEbiography}[{\includegraphics[width=1in,height=1.25in,clip,keepaspectratio]{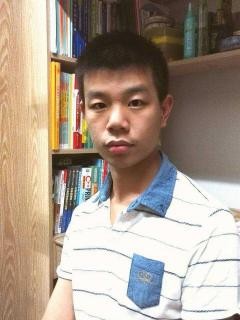}}]{Darong Huang}
is currently a Ph.D. candidate in Electrical Engineering (EE) in the Embedded Systems Laboratory (ESL) at Swiss Federal Institute of Technology Lausanne (EPFL). He received his B.Sc. and M.Sc. degrees in Electrical Engineering from the University of Electronic Science and Technology of China in 2016 and 2019, respectively. His research interests focus on the thermal and reliability management of microprocessors.
\end{IEEEbiography}
\vskip -1.4\baselineskip plus -1fil
\begin{IEEEbiography}%
  [{\includegraphics[width=1in,height=1.25in,clip,keepaspectratio]{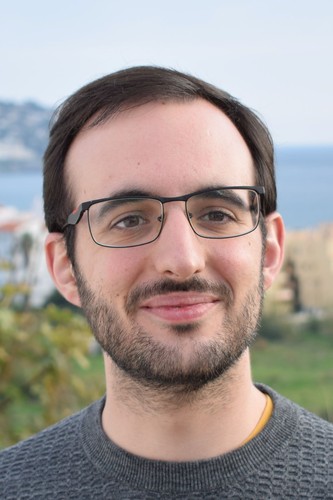}}]
  {Luis Costero}  Luis M. Costero is an assistant professor in the Complutense University of Madrid (UCM), Spain since 2022. He was a post-doctoral researcher in the Embedded Systems Laboratory (ESL) at Swiss Federal Institute of Technology in Lausanne (EPFL) until Feb. 2022. He received his Ph.D. degree in Computer Engineering from the Complutense University of Madrid (UCM) in 2021. He also obtained a M.Sc. in Computer Engineering from the same university. His main research areas involve high performance computing, asymmetric processors, power consumption and real-time resource management.
\end{IEEEbiography}
\vskip -1.4\baselineskip plus -1fil
\begin{IEEEbiography}[{\includegraphics[width=1in,height=1.2in,clip,keepaspectratio]{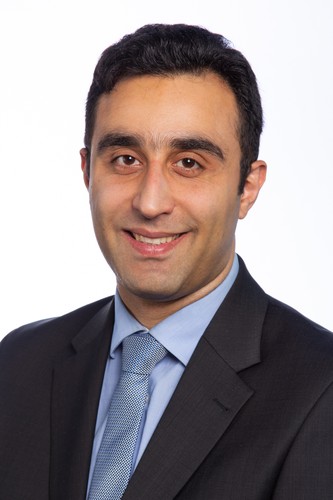}}]{Ali Pahlevan}
is a post-doctoral researcher in the Embedded Systems Laboratory (ESL) at Swiss Federal Institute of Technology Lausanne (EPFL). He received his Ph.D. degree in Electrical Engineering (EE) from EPFL in 2019, M.Sc. degree in Computer Engineering from Sharif University of Technology (SUT) in 2012, and B.Sc. degree in Computer Engineering from Ferdowsi University of Mashhad (FUM) in 2010.
He has published over 15 research papers in top international journals and conferences.
\end{IEEEbiography}

\vskip -1.4\baselineskip
\begin{IEEEbiography}[{\includegraphics[width=1in,height=1.25in,clip,keepaspectratio]{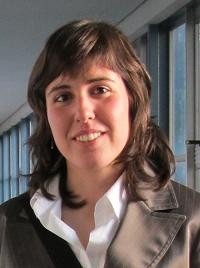}}]{Marina Zapater} 
is associate professor in the School of Engineering and Management of Vaud (HEIG-VD) at the University of Applied Sciences Western Switzerland (HES-SO) since 2020. She was a Postdoctoral Research Associate with the Embedded System Laboratory (ESL) at EPFL until February 2020, and is currently an external collaborator of ESL-EPFL. She received her Ph.D. degree in electronic engineering from UPM, Spain, in 2015. Her research interests include thermal, power and performance design and optimization of heterogeneous architectures, from edge devices to high-performance computing processors; and energy efficiency in servers and datacenters. In these fields, she has co-authored more than 75 papers in top-notch conferences and journals. She is an IEEE and CEDA member, and CEDA Assistant VP of finance (2019-2020).
\end{IEEEbiography}
\vskip -1.4\baselineskip
\begin{IEEEbiography}[{\includegraphics[width=1in,height=1.25in,clip,keepaspectratio]{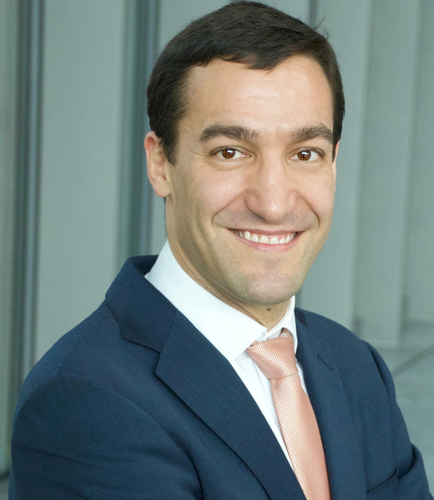}}]{David Atienza} (M'05-SM'13-F'16) is professor of electrical and computer engineering, and head of the Embedded Systems Laboratory (ESL) at EPFL, Switzerland. He received his PhD in computer science and engineering from UCM, Spain, and IMEC, Belgium, in 2005. His research interests include system-level design methodologies for high-performance multi-processor system-on-chip and low power Internet-of-Things systems, including thermal-aware design for MPSoCs and many-core servers, and edge AI architectures for wearables and IoT systems. He is a co-author of more than 350 papers, one book and 14 patents. Among others, Dr. Atienza has received the ICCAD 2020 10-Year Retrospective Most Influential Paper Award, the 2018 DAC Under-40 Innovators Award, and an ERC Consolidator Grant in 2016. He is an IEEE Fellow and an ACM Distinguished Member.
\end{IEEEbiography}

\vfill